CHAPMAN UNIVERSITY | INSTITUTE FOR QUANTUM STUDIES

**REGULAR PAPER**

Check for updates

# Nashian game theory is incompatible with quantum physics

**Michal Baczyk · Ghislain Fourny** [ORCID]

**Abstract** We suggest to look at quantum measurement outcomes not through the lens of probability theory, but instead through decision theory. We introduce an original game-theoretical framework, model and algorithmic procedure where measurement scenarios are multiplayer games with a structure all observers agree on. Measurement axes and, newly, measurement outcomes are modeled as decisions with nature being an action-minimizing economic agent. We translate physical notions of causality, correlation, counterfactuals, and contextuality to particular aspects of game theory. We investigate the causal consistency of dynamic games with imperfect information from the quantum perspective and conclude that counterfactual dependencies should be distinguished from causation and correlation as a separate phenomenon of its own. Most significantly, we observe that game theory based on Nash equilibria stands in contradiction with a violation of Bell inequalities. Hence, we propose that quantum physics should be analyzed with non-Nashian game theory, the inner workings of which we demonstrate using our proposed model.

## 1 Introduction

The interrelation between quantum theory and game theory was established in a seminal paper by Eisert, Wilkens, and Lewenstein [1]. This work introduced the quantization of nonzero sum games and showed that it is possible to design quantum strategies acting as unitaries on an input system that, through entanglement, generally outperform classical strategies. The embedding of quantum physics into games, via additional primitives that the agents have at their disposal, became a new research framework [2–4]. Exploring its capabilities led to novel ideas, such as "coherent equilibria" [5] and created links to other disciplines like quantum information theory and quantum algorithms [6], market analysis [7], and many others [8,9].

In this paper, we investigate a yet uncharted research direction that is complementary to the mentioned studies—we go the opposite way and transfer the game-theoretic concepts to the quantum domain. Note in particular that, unlike in quantum games, we consider game theory with only classical (not quantum) strategies, in order to show that recent developments in game theory bring innovative insights that contribute to understanding the defining

M. Baczyk · G. Fourny (✉)
Department of Computer Science, ETH Zürich, Universitätstrasse 8, 8092 Zurich, Switzerland
e-mail: ghislain.fourny@inf.ethz.ch

Springer

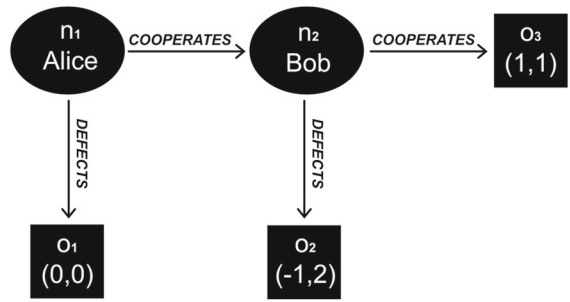

**Fig. 1** The promise game. Each box with the name of the player constitutes a decision node. The edges lead to possible future scenarios and eventually to outcomes of the game, which are the leaves in the decision tree. Each leaf shows the payoffs of the players—the number on the left being Alice's payoff and on the right Bob's payoff. Each resolution of the game is a path going from the root to an outcome

aspects of quantum theory. We suggest to model measurement outcomes as decisions by action-minimizing nature, playing against, and symmetric to, the choice of measurement axis by utility-maximizing physicists.

We assume the reader may not be familiar with game theory and start by highlighting the fundamental concepts and structures of game theory with dynamic games (e.g., the promise game) and strategic games (e.g., the prisoners' dilemma). The two paradigms represent players, outcomes and payoffs similarly:

- a set $\mathcal{P} = \{1, \ldots, N\}$ of unique players (we will also use first names and their initials for convenience);
- a set $\mathcal{Z}$ of possible outcomes of the game;
- a family of payoff functions $(u_i)_{i \in \mathcal{P}}$, one for each player, from $\mathcal{Z} \to \mathbb{R}$. They determine the reward $u_i(o)$ that each player $i$ gets given the outcome $o$ of the game.

Dynamic and strategic games then differ in how outcomes are structured.

Figure 1 shows the promise game, which is a dynamic game [10–12]: agents make their moves in turns, i.e., the consecutive choices are time-like separated events. In this scenario, we have $\mathcal{P} = \{1, 2\}$, $\mathcal{Z} = \{o_1, o_2, o_3\}$ and, for example, $u_1(o_2) = -1$. The outcomes of a dynamic game are organized as the leaves of a tree structure, where players make a decision at each intermediate node, going down the tree and eventually reaching one of the outcomes. Formally, the structure consists of:

- a set $\mathcal{H}$ of decision nodes, with each node $N \in \mathcal{H}$ associated with a player $\rho(N) \in \mathcal{P}$;
- a set of actions $\mathcal{A}$, with each node $N \in \mathcal{H}$ associated with a subset of actions $\chi(N) \subseteq \mathcal{A}$ that the player $\rho(N)$ can pick from;
- a successor function $\sigma$ mapping each node $N \in \mathcal{H}$ and action $a \in \chi(N)$ to the next node $\sigma(N, a) \in \mathcal{H} \cup \mathcal{Z}$, either a decision node or outcome. $\sigma(N, a)$ is undefined for $a \notin \chi(N)$.

Here $\mathcal{H} = \{n_1, n_2\}$, $\rho(n_1) =$ Alice, $\rho(n_2) =$ Bob, $\mathcal{A} = \chi(n_1) = \chi(n_2) = \{$cooperates, defects$\}$, $\sigma(n_1, \text{cooperates}) = n_2$, $\sigma(n_1, \text{defects}) = o_1$, $\sigma(n_2, \text{cooperates}) = o_3$, $\sigma(n_2, \text{defects}) = o_2$. We illustrate this game by the following example. Alice, the baker, starts and decides ($n_1$) whether or not to give a loaf of bread to Bob. Afterward, Bob, if given the bread, chooses ($n_2$) whether or not to pay Alice for it. In the Nash paradigm, it is rational for Bob not to pay for the bread because $u_{\text{Bob}}(o_2) > u_{\text{Bob}}(o_3)$. Likewise, Alice, anticipating this, does not hand over the bread because $u_{\text{Alice}}(o_1) > u_{\text{Alice}}(o_2)$.

The promise game is a game with *perfect information*, since Bob acquires the information about Alice's choice. If Bob were presented with an opaque bag either containing or not containing the bread then it would be a game with *imperfect information*. In a game with imperfect information, players may not be fully informed about decisions made at ancestor nodes; this is modeled with equivalence classes of decision nodes, called information sets, that make them indistinguishable. In this work, in diagrams presenting games with imperfect information, information sets are denoted with dashed lines in the dynamic game structure.

| ALICE ╲ BOB | COOPERATE | DEFECT |
|---|---|---|
| COOPERATE | 2,2 | 0,3 |
| DEFECT | 3,0 | 1,1 |

**Fig. 2** The prisoners' dilemma. The game diagram possesses a matrix rather than a tree structure, since for this game the agents are isolated from each other and do not make their moves in turns. The resolution of the game is a single entry in the table indicating the payoffs for the players in that game realization—the number on the left being Alice's payoff and on the right Bob's payoff

Figure 2 shows the prisoners' dilemma, which is a strategic game [12–14]: agents choose their strategies in isolation from each other. Although it is not necessarily required, this can be thought of as decisions being space-like separated. What is sufficient is to ensure with other means that there is no communication between the agents.

Strategic games organize their outcomes in a different way than dynamic games, namely as a matrix indexed by the agents' strategies. Each player $i$ can pick a strategy from their set of strategies $\mathcal{S}_i$. The set of outcomes $\mathcal{Z}$ is then the Cartesian product of all strategies: $\mathcal{Z} = \prod_{i \in P} \mathcal{S}_i$. Each outcome is thus also called a strategy profile.

In our example, $\mathcal{P} = \{\text{Alice}, \text{Bob}\}$, $\mathcal{S}_1 = \mathcal{S}_2 = \{\text{defect}, \text{cooperate}\}$, the outcomes are $\mathcal{Z} = \mathcal{S}_1 \times \mathcal{S}_2 = \{(\text{cooperate}, \text{cooperate}), (\text{cooperate}, \text{defect}), (\text{defect}, \text{cooperate}), (\text{defect}, \text{defect})\}$, and for example $u_1(\text{cooperate}, \text{defect}) = 0$. This models the situation in which two burglars caught by the police are put in separate jail cells and are interrogated. If a burglar snitches on their teammate, they defect and if they stay silent, they cooperate. In the Nash paradigm, it is rational for each agent to defect because treating the other agent's decision as fixed (no matter what it is), the choice to defect always gives a higher payoff. However, it is not the most rewarding outcome for both Alice and Bob.

Formally, in a strategic game, a Nash equilibrium [12,15] $E$ is an outcome jointly reached by all the players such that, if any player *had chosen* (mind the subjunctive conditional) another strategy assuming that others strategies are fixed (this is called a *unilateral deviation*), then it would have resulted in a lower payoff for them. Formally, $E = (s_1^*, s_2^*, \ldots, s_N^*) \in \mathcal{Z}$ is a Nash equilibrium if

$$\forall i \in \mathcal{P} \quad \forall s_i \in \mathcal{S}_i : \quad u_i(s_1^*, s_2^*, \ldots, s_i^*, \ldots, s_N^*) \geq u_i(s_1^*, s_2^*, \ldots, s_{i-1}^*, s_i, s_{i+1}^*, \ldots, s_N^*)$$

Under the Nashian paradigm, there is a procedure to convert any dynamic game to a strategic game with the same outcomes called its strategic form. The choices, for each player, are "flattened" to a unique set of strategies that corresponds to a combination of choices the player makes in each one of their nodes; this can be seen seen as some sort of a (non-contextual) masterplan that players can come up with ahead of the dynamic game, anticipating any possible play. This naturally extends Nash equilibria to a dynamic game: these are the Nash equilibria of its strategic form. A particular instantiation of a Nash equilibrium in a dynamic game is the Subgame Perfect Equilibrium [16], which is obtained by a backward induction starting at the leaves and optimizing choices all the way up to the initial node.

The two categories of games above have been shown to be special cases of a more general category of games, which are the focus of this paper, called spacetime games with perfect information [17]. In spacetime games, players can make decisions at any location in Minkowski spacetime, which is the flat version of spacetime with no gravitational effects. Dynamic games are spacetime games with timelike-separated choices. Strategic games are spacetime games with space-like separated choices. The structures organizing the outcomes of a spacetime games are summarized in the Methods section. An algorithm to convert any spacetime game with perfect information to a dynamic game with imperfect information, as well as to a strategic form is given in [17]. In consequence, Nash equilibria also naturally extend to spacetime games.

A Nash equilibrium considers unilateral deviations only, which directly corresponds to assuming that the agents have free choice in quantum foundations. However, this Nashian free choice is not the only approach to game theory. A whole spectrum of alternate, non-Nashian approaches exists, such as the Perfect Prediction Equilibrium [18], the Translucent Equilibrium [19], the Perfectly Transparent Equilibrium [20,21], Superrationality [22], and Minimax-Rationalizability [23]. For some games, there exist non-Nashian resolutions with potentially higher payoffs for all the agents than in the Nashian framework. In our examples, the Nash equilibrium fails to reach Pareto-optimal outcomes: (1,1) in the promise game and (2,2) in the prisoners' dilemma game. $o_3 = (1, 1)$ is an equilibrium in the promise game in the Perfect Prediction Equilibrium framework because of the following reasoning. If Bob were to pick $o_2$ rather than $o_3$, Alice would have forecast it as the rational thing for him to pick and she would have chosen $o_1$. But this would be contradictory with him playing at $n_2$ at all. This is a reductio-ad-absurdum proof that, if he was given, by rational Alice, the possibility to play at $n_2$, it logically follows that he picks $o_3$ to obtain a consistent timeline. We provide a detailed description in the Methods section.

In this work, we use the framework of spacetime games with perfect information [17] (a concise summary thereof is given in the Methods section) which can model any intricate scenario combining the cases of decisions being both space-like and time-like separated and for which there is a well-defined Perfectly Transparent Equilibrium, which is at most unique and Pareto-optimal if there are no ties in the payoffs ("generic position"). In the Results section, first, we propose a framework in which we establish the notions of counterfactual dependency/independence and Nashian as well as non-Nashian free choice. Second, we analyze the Bell experiment setup as a game, hence demonstrating the inner workings of an algorithm allowing to translate a physical scenario involving quantum measurements into a game, where the decision nodes are both choices of measurement axes by physicists and choices of measurement outcomes by nature, and where nature's payoff maximization can be understood through the glasses of the principle of least action. Furthermore, we also describe in details how the optimal strategy resolutions to this game are found using both Nashian and non-Nashian assumption. Not only is the non-Nashian approach consistent with violation of Bell inequalities; by construction, it also allows for contextuality and passage of time.

## 2 Results

### 2.1 The framework

We propose a mathematical framework, agnostic to the free choice assumption and consistent with game theory; our framework is capable of describing any quantum experiment scenario in which experimental setup options are selected, the results of a measurement process are obtained, and such experiments are carried out in arbitrary numbers and at arbitrary positions in spacetime. For that, we consider an input–output parameters model akin to that introduced by Wharton et al [24], where input parameters reflect the chosen characteristics of the apparatus and output parameters correspond to measurements outcomes. However, we extend it to a more general framework in which input and output parameters need not all be jointly defined in every history, and might thus be contextual. Additionally, this more general framework allows to introduce a novel notion of symmetry between input and output parameters. The model presented in this paper is compatible with quantum causal models [25–27] and more generally with the process matrix framework [28]. Indeed, an information–theoretic causal structure is naturally present in these frameworks, which can be interpreted as relativistic causality, leading naturally to a spacetime game in which the nodes correspond to the input and output parameters of a particular process matrix model.

To start, we define the following notions:

**Multihistory** The multihistory $\Omega$ is the set of all possible histories of the quantum experiment scenarios. Each $\omega \in \Omega$ represents one way the history could be and is endowed with a Minkowski spacetime $\mathbb{R}^4$. $\Omega$ can be endowed with a $\sigma$-algebra $\mathcal{F}$, making $(\Omega, \mathcal{F})$ a measurable space. $\Omega$ can also be endowed with a probability measure $P$.

In each history, there are parameters, located in spacetime, that can take values or be undefined. A history can, thus, alternatively be seen as a history of parameter values.

**Parameter** An (input/output) parameter $N$ is a random variable on $\Omega$ with values in some target set $\chi(N) \cup \{\bot\}$. The special value $\bot$ is an additional element of this target set that expresses the possibility of $N$ being undefined in a history $\omega$, i.e., $N(\omega) = \bot$. Furthermore, we require that each parameter has a well-defined position in Minkowski spacetime—the same in all histories.

While parameters are modeled as random variables on $\Omega$ (which are merely functions from $\Omega$ to a target set), we note that this does not imply that their values are selected at random. Our model allows an interpretation of the parameters as being selected at random with some probability measure on $\Omega$, which is the common interpretation of quantum physics, but it also allows an interpretation in which the values of the parameters are selected through a decision model, i.e., in which the parameters are nodes in a game and the probability measure can be interpreted subjectively, but with a different assumption of free choice.

Let $\mathcal{H}$ denote the set of all (input and output) parameters in a given model and let $\rho$ be a function that maps each decision $N \in \mathcal{H}$ to the agent $\rho(N) \in \mathcal{P}$ who makes this decision and picks a value in $\chi(N)$. Parameters can model all the macroscopic decision processes leading to a numerical or categorical result. Note that game theory allows, additionally to humans, for nature to be a player [29], which typically is implemented as a random number generator. Our framework also allows moves by nature (output parameters), but does not restrict them to be necessarily made with a random number generator.

## 2.2 Counterfactuals

Next, we define formally the concept of *counterfactual dependency* in the proposed framework. Counterfactual dependencies [30, 31] are needed when we consider alternate values parameters could have taken, and what alternate history this would have corresponded to. In plain English, they can be expressed as subjective conditionals, such as: "if Bob had measured the spin along the $z$ axis, then the outcome would have been 1". We already encountered such statements in the Introduction where we considered deviations from the (Nash or non-Nashian) equilibrium. To give counterfactual dependencies formal meaning, we begin with employing the notion of closest history in which a fact is true, which is utilized in the theory of counterfactuals [31] and in game theory [23]:

**Closest history in which a fact is true** Given a history $\omega \in \Omega$, a parameter $N$ (which may or may not be defined in $\omega$) and a value $a \in \chi(N)$ that $N$ can take, we assume (as an axiom called "uniqueness" in the theory of counterfactuals [30, 31]) that there is exactly one closest history in which $N = a$ is true, and denote this history $f_{N=a}(\omega)$.

Let us consider some fact that is false in the current history, the actual world. Intuitively, the closest history in which this fact is true has an alternate past that causes this fact to be true, with all other facts changed by a minimal amount. For example, one could ask what minimal set of events would had had to be different so that the president of the United States be Arnold Schwarzenegger. This is a world as similar as possible to ours, but in which the chain of events caused Arnold Schwarzenegger to be elected president.

Then, the counterfactual dependency between two (input or output) parameters of the quantum system can thus be formally defined as follows:

**Counterfactual dependency** Given a history $\omega \in \Omega$, two parameters $N$ and $M$ (which may or may not be defined in $\omega$) and two values $a \in \chi(N)$ and $b \in \chi(M)$ that $N$ and $M$ can take, $N = a$ is said to *counterfactually imply $M = b$* in $\omega$, which is denoted $N = a >_{\omega} M = b$, if $M = b$ is true in $f_{N=a}(\omega)$.

This definition puts no restriction on the spacetime location of counterfactually dependent variables, in particular, in general, a spacetime variable may be counterfactually dependent on another located in its future light cone, or on another that is spacelike-separated. In this sense, it is agnostic to the possible views on free choice that we will soon discuss (which includes future-input dependent models).

We also define the notion of counterfactual independence of A from B.

**Counterfactual independence** Given two parameters $N$ and $M$, $N$ is said to be *counterfactually independent from $M$* if for all worlds $\omega \in \Omega$,

$$N(\omega) \neq \perp \wedge M(\omega) \neq \perp \implies \forall b \in \chi(M), M = b >_{\omega} N = N(\omega)$$

Counterfactual independence means that, in any history $\omega$ in which both parameters are defined, in the closest history $f_{M=b}(\omega)$ in which $M$ takes a (possibly different) value $b$, $N$ has the same value as in $\omega$.

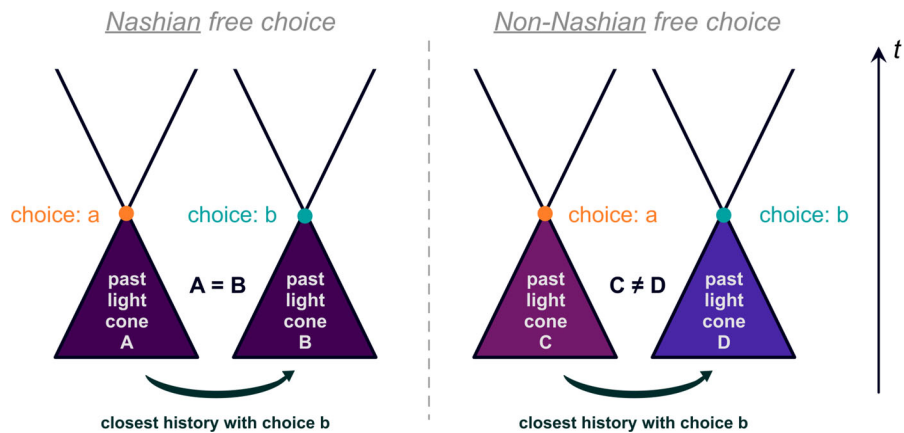

**Fig. 3** Nashian free choice vs. non-Nashian free choice. The difference between the two concepts of free choice is illustrated with spacetime diagrams, with time on the vertical axis and space on the horizontal axis. In both cases, the actual world is shown on the left. In the actual world, choice a is made. On the right, for each paradigm, the closest world in which choice b is made is shown. Under Nashian free choice, commonly assumed in the quantum literature, the closest world in which choice b is made has the exact same past light cone, which is another way to say that the past is counterfactually independent of the choice. On the other hand, under non-Nashian free choice, the closest world in which choice b is made has a different past light cone, which is another way to say that, if the choice had been different, the past leading to this choice would have been different, too. This is a counterfactual dependency and it should not be confused with a causal dependency: it is the past light cone that jointly causes choice a (or b), respecting the flow of time, and not the other way round. In other words, under non-Nashian free choice, if the choice had been different, then this different choice should have been caused by a different past chain of events leading to it

## 2.3 Free choice

The quantum foundations literature commonly assumes that agents carrying out quantum experiments freely choose their measurement axes. This is typically done by assuming statistical independence from anything not in the future light cone. However, we previously made an argument [17] that statistical independence implies counterfactual independence in the spirit of David Lewis's work [31]; and a second argument could be made that statistical independence without counterfactual independence would require the undesirable feature of fine-tuning [27]. Thus, we now reformulate this notion of free choice, which we call Nashian free choice, based on counterfactuals:

**Nashian free choice** A parameter $N$ is said to be *freely chosen in the Nash sense* if any parameter $M$ that is not in its future light cone is counterfactually independent from $N$. (This definition derives from [32], where mathematical formulation of the notion of free choice was introduced.)

**Non-Nashian free choice** A parameter $N$ is said to be *freely chosen in the non-Nashian sense* if for every $a \in \chi(N)$ there exists a history $\omega$ in which $N(\omega) = a$. Non-Nashian free choice is also called full support in the quantum foundations literature [32].

A diagram contrasting the differences between Nashian and non-Nashian free choice is presented in Fig. 3. Consider an event in Minkowski spacetime which corresponds to an agent making a decision, in some history $\omega$. This decision can be modeled as a parameter $X$ (say, $X(\omega) = a \in \chi(X)$) and the past light cone (including the boundary but not $X$ itself) of $X$ can be treated, from the game theory perspective, as a history of all the past parameters' values. Accordingly, if we assume $X$ is freely chosen in the Nashian sense, then in the closest history in which $X = b$ (closest with respect to $\omega$) for some $b \in \chi(X) \setminus \{a\}$, the past light cone of $X$ (the region of spacetime that all observers agree precedes $X$) is the same. That is not the case for the non-Nashian framework in which the past light cone of $X$ is counterfactually dependent on the value of $X$.

At that place, we would like to model a Bell-type experiment from the game theory point of view. Consider a game with four players $P = \{1(\text{Alice}), 2(\text{Bob}), 3(\text{Carol}), 4(\text{David})\}$. Alice and Bob are experimental physicists sharing an entangled pair of particles and deciding which measurement settings are used for the experiment, i.e. they determine the input parameters of the system. Alice can choose from $\{a_1, a_2\}$ and Bob from $\{b_1, b_2\}$. Carol's

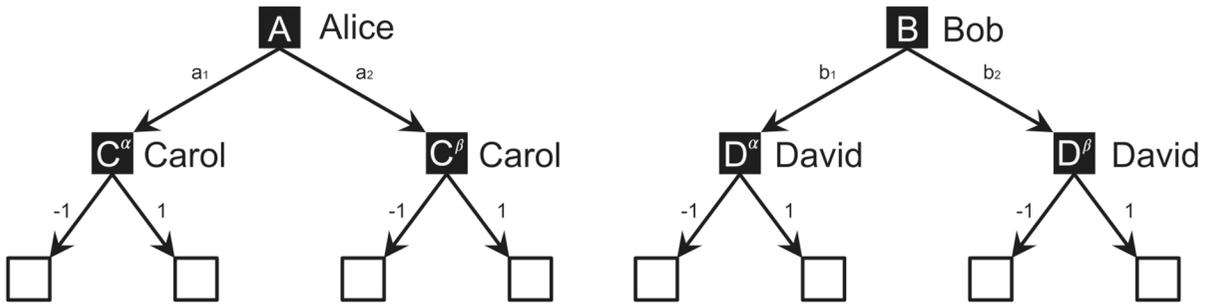

**Fig. 4** The causal structure for a Bell-type experiment, as a spacetime game but with the payoffs not shown. It is a Directed Acyclic Graph where nodes represent (input/output) parameters and edges correspond to future-directed time-like separation for the relevant pairs of parameters. For readability, only the irreflexive and transitive reduction of the causal dependencies is shown. In that setting six input–output parameters are specified with the following map from moments of decisions (parameters) to players: $\rho(A) =$ Alice, $\rho(B) =$ Bob, $\rho(C^\alpha) = \rho(C^\beta) =$ Carol, $\rho(D^\alpha) = \rho(D^\beta) =$ David. Agents' payoffs are left open. By specifying payoffs for each possible outcome, a spacetime game is obtained

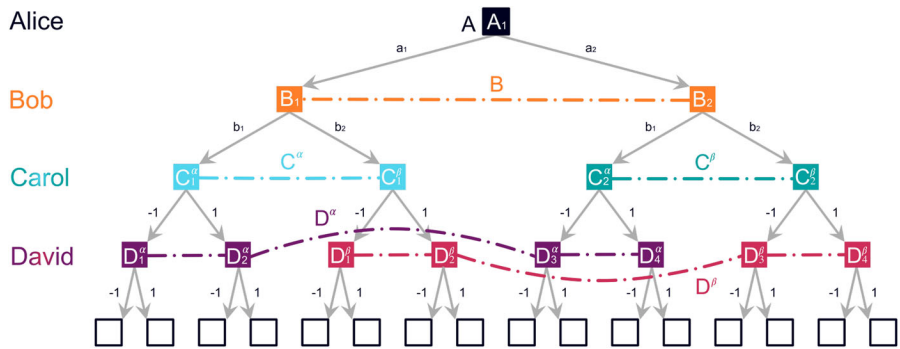

**Fig. 5** The conversion of the Bell-type experiment to a dynamic game with imperfect information. Each level of a tree corresponds to a different player, from top: Alice, Bob, Carol and David. The dotted lines connect the decision nodes that come from the same moment of decision presented in Fig. 4. The payoffs were picked arbitrarily and are parameters to this model

and David's choices model the measurement outcomes (output parameters) which, for both of them, can be either $-1$ or $1$. Although Carol and David do not represent human players, from the perspective of game theory they can be treated as agents that emulate what happens to the particles that are being measured. If we assume Nashian free choice, then because of the Free Will Theorem (Conway, Kochen [33,34]), Carol and David posses Nashian free choice as Alice and Bob do.

We further impose that both events corresponding to Alice's and Carol's choices are space-like separated from both events representing Bob's and David's decisions. Hence, the causal structure of this setup can be depicted as in Fig. 4. Note that we explicitly distinguish Carol's choice in the history for which Alice chooses $a_1$ from her decision in the history for which Alice chooses $a_2$ by modeling these two decisions with two different parameters $C^\alpha$ and $C^\beta$. The same for David. In the language used in [17] (please see the Methods section for more details), we say that the *contingency coordinate* for $C^\alpha$ to be defined is $A = a_1$ (similarly for other output parameters).

Note that, while our examples involve binary decisions, this is not a limitation of the framework and it is possible to have any number of choices for each information set.

### 2.4 Conversion to a game

With the spacetime structure of the game set, parameters and the contingency dependencies between them defined, we can use the algorithm established in [17] to convert the above described scenario to a dynamic game with imperfect information (see Fig. 5). First, let us emphasize that the structure of the obtained game is agreed on by all observers, since the space-like and time-like separation relations of the decision points are invariant under Lorentz transformations (which describe the difference between what observers see and when). Secondly, let us observe that the game can be seen as a process in which agents make decisions in turns. However, a crucial aspect must be added to this view, namely that agents making a decision do not know in which history they are located. They only know their decision points, which are equivalence classes formed on the set of decision points. This notation is formally captured with information sets in games with imperfect information.

Having constructed a game representing a Bell-type experiment, we observe that the game-theoretic formulation by its very nature imposes the locality and realism assumptions. The reason is that decisions made by agents can be modeled by functions that represent their reasoning processes and the information available to each player when making a move is contained in their past light cone. Locality is thus enforced by the information sets, which restrict the available information to that coming from the past light cone. Realism is enforced by the modeling of players' strategies, which map some or all of the information sets to one of the actions available at these information sets. Hence, because of the violation of Bell inequalities, if the game theoretic and quantum perspectives are to be reconciled, the assumption of free choice in the Nash sense must be weakened. It means that Nashian game theory is intrinsically inconsistent with quantum physics and other approaches to game theory should be considered. In the following, we present the characteristics of the non-Nashian approach and how it additionally captures the defining quantum features such as contextuality.

Although Nashian game theory would contradict a violation of Bell inequalities, to demonstrate further differences between Nashian- and non-Nashian (Perfectly Transparent Equilibrium [21])-free choice assumptions, let us consider how the presented Bell-type game is solved for an equilibrium in both cases. We introduce some concrete payoffs, whereby the choice of payoff structure is a free parameter of each model instantiated from our framework (i.e., if a principle of least action is to be used for nature's payoffs, it is left open which action is minimized).

In Fig. 6, we see a resolution of the game in the Nashian framework (with arbitrarily chosen payoffs): $A = a_1$, $B = b_1$, $C^\alpha = -1$, $C^\beta = -1$, $D^\alpha = 1$, $D^\beta = -1$. The equilibrium found in such a manner is stable against unilateral deviations. For example, if Alice had chosen $A = a_2$ instead of $A = a_1$, keeping the other players' choices fixed it would have led to the outcome $(9, 4, 1, 14)$ in which she would have obtained 9 which is no more than 10. The same reasoning applies to all other players and decisions. Note that, in the Nash paradigm, the choices have to be specified for all the decision nodes including unreached ones (here $C^\beta$ and $D^\beta$), and an assignment of choices is called a Nash strategy. This feature corresponds directly to non-contextuality which is excluded from quantum premise by means of the Kochen–Specker theorem [35,36].

## 2.5 Resolution of the game

We now analyze how the same game is solved when the non-Nashian form of free choice is assumed, again with arbitrarily chosen payoffs. Figure 7 presents the forward induction process which leads to the Perfectly Transparent Equilibrium solution. The principles utilized in the elimination procedure correspond to the ones introduced in the Methods section for the Perfect Prediction Equilibrium framework. We start the analysis by realizing that Alice and Bob have to take at least one decision in all the possible histories. We do not consider yet the choices of Carol and David because their opportunities to make a decision or not depend on what Alice and Bob do before them. That is why in Fig. 7a Alice and Bob are put in bold frames to emphasize that we firstly examine the situation from their point of view.

Let us observe that the minimal payoff that Alice could get if she chose $a_1$ is 6, while it is 1 for $a_2$. Her so-called "maximin" is thus 6; it is the maximal enforceable payoff for Alice, and she has a guarantee to get at least this by choosing $a_1$. The minimal payoff that Bob could get if he chose $b_1$ is 2, while it is 1 for $b_2$. His so-called "maximin" is thus 2. Hence, from the first principle introduced in the Methods paragraph explaining the Perfect Prediction Equilibrium concept, we conclude that the outcomes giving Alice a payoff of 1, 2, 3, 4, or 5 (less than 6) in the $a_2$ subtree are discarded for the remainder of the reasoning: If any of these outcomes had been the solution, Alice would have deviated to $a_1$ because this would have guaranteed her a payoff of at least 6, which leads to a contradiction

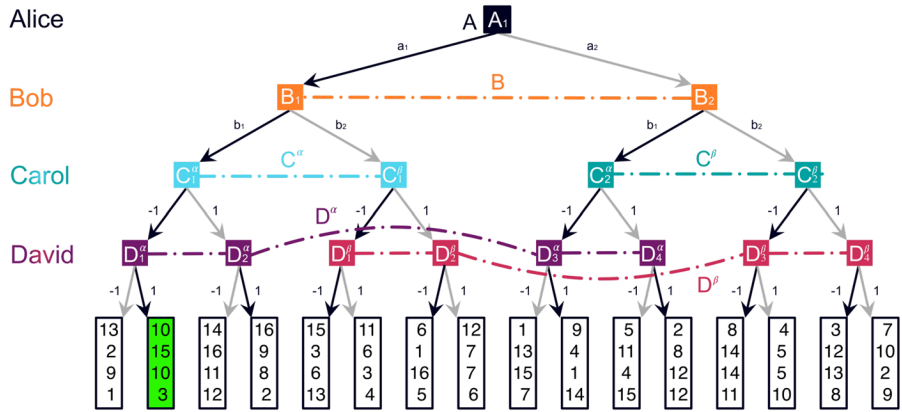

**Fig. 6** The Nash equilibrium resolution of a Bell-type scenario. We assign payoffs to each of the outcomes in a form of number tuples for which the first number is the Alice's, the second Bob's, the third Carol' and the fourth David's payoff. Black arrows show which choices the agents make. Note that, in a Nash equilibrium, decisions are also defined for unreached nodes, which is a non-contextual feature. The resolution of the game is a path following black arrows from the root of the tree to one of the leaves. The found equilibrium is (10, 15, 10, 3)

(Grandfather's paradox). Likewise, outcome (6, 1, 16, 16) is eliminated using the same principle but considering Bob's decision. If it had been the solution, Bob would have deviated to $b_1$ because this would have guaranteed him a payoff of at least 2, leading to a contradiction (Grandfather's paradox).

Now (see 7a), the maximin payoff for Alice is 10 (for $a_1$) and the maximin payoff for Bob is 3 (for $b_2$). Thus, we eliminate four more outcomes. Once this is done, the right branch is empty (see 7b). Hence, we know that, in the solution, Alice chooses $a_1$. We can thus add to considerations Carol's perspective since we know that the contingency coordinates of her decision point are met (second principle). That is why we put Carol in bold frame.

The third round of elimination, once again based on the first principle is depicted in subfigure 7c. Bob's maximin is 9, and Carol's maximin is 7. Three more outcomes are discarded; we see that Bob chooses $b_1$ and that now also David gets to choose.

In the fourth and last round, Carol's maximin is 10 and David's maximin is 12. Outcome (16, 9, 8, 2) is eliminated by Carol (using the first principle) and she chooses 1 because it gives her higher payoff (second principle). Then Daniel chooses (14, 16, 11, 12) as the only logical possibility left. This outcome is the Perfectly Transparent Equilibrium for the considered game.

It is important to point out that contrary to the Nashian case, decisions are only assigned on a path leading to the equilibrium outcome. Interpreting this fact in a physics context we can say that only the events happening in the actual history are elements of reality. Additionally to that we can conclude that the resolution for the Perfectly Transparent Equilibrium concept is inherently contextual. Finally, a forward induction reasoning is computationally more efficient than the backward induction used in Nash game theory and more importantly it possesses a very different physical interpretation. Backward induction requires starting from every single possible future and in a forward induction, only decisions on the actual equilibrium path history ever get assigned, and this happens in a dynamics going top-down, which allows a partial resolution of a game up to a specific period of time, pausing, and then resuming at will. The algorithm used to solve for non-Nashian equilibrium, hence, resembles a passage and directionality of time.

## 3 Discussion

As a necessary step for defining the Nashian and non-Nashian free choice, we have defined *counterfactual dependencies*. We would like to compare this concept first with causation and second with correlation. Hence, we aim to show that counterfactual dependency has further importance than being a piece of terminology.

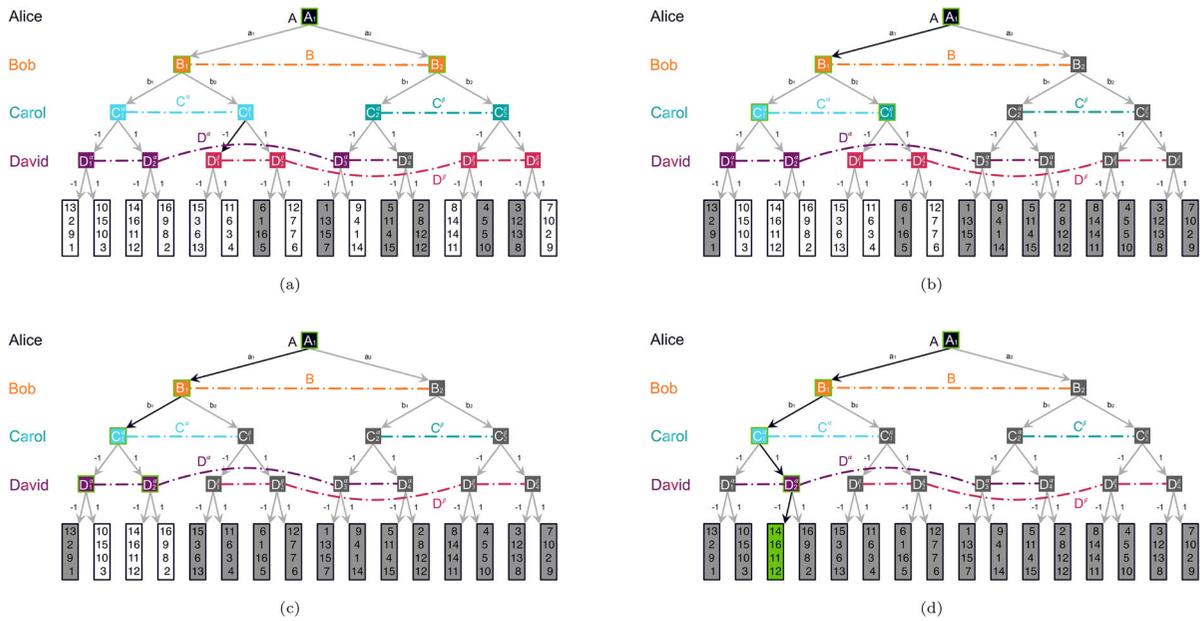

**Fig. 7** A non-Nashian resolution (in this case the Perfectly Transparent Equilibrium) of the Bell-type scenario. Subfigures **a**–**d** show consecutive steps in a forward induction procedure. To discard some outcomes, possible decision choices of agents inscribed in bold frames are considered. The eliminated outcomes are marked with dark gray color and are no longer considered. We assign payoffs to each of the outcomes in a form of number tuples for which the first number is the Alice's, the second Bob's, the third Carol' and the fourth David's payoff. Black arrows show which choices the agents make. Note that, in the Perfectly Transparent Equilibrium, decisions are undefined at unreached nodes. The resolution of the game is a path following black arrows from the root of the tree to one of the leaves. The found equilibrium is (14, 16, 11, 12)

For the purpose of our discussion, we define causation in terms of special relativity: a parameter $A$ defined is said to be *causally dependent* on another parameter $B$ if $B$ is in the past light cone (including the boundary but not the tip) of $A$. (Note that this statement has a physically meaningful interpretation only in the histories in which both $A$ and $B$ are defined.)

Distinguishing the notion of counterfactual dependency as a separate construct becomes instrumental when analyzing a Bell-type experiment. Suppose Alice and Bob, who share an entangled pair of particles, perform measurements that are space-like separated, whose outcomes are either $-1$ or $+1$. They obtain opposite results $A = -B$ ($A$ for Alice, $B$ for Bob). Parameter $A$ is space-like separated from parameter $B$, hence it is causally independent from $B$ – no need to invoke cause and effect. However, $A$ is counterfactually dependent from $B$ because if Bob had measured a different value, Alice would have measured a different value as well.

As for correlations, in our framework, they correspond to the use of statistics on the values of input and output parameters that actually happen and get collected on some record medium within the one and same history.

At this point, as a main takeaway of the paper, we would like to address the fact that Nashian game theory, as established in the Results section, is incompatible with violation of Bell inequalities. If we were to stay in the Nashian free choice framework, it would be possible, from the game theory perspective, to eliminate the obtained contradiction by introducing chance moves for all the agents [29] (agents' decisions governed by random processes). This approach, although it resolves the raised issue, puts tight restrictions for research at the intersection of game theory and quantum theory because most of the game-theoretic concepts are not applicable to scenarios evolving in a random way: game theory, and the concept of rationality, is useful only if a game contains at least some moves that are not chance moves. A game with only chance moves is better modeled with concepts from the theory of Bayesian networks [37].

That is why, we would like to propose to circumvent the brought up incompatibility by replacing the Nashian-free choice with the non-Nashian free choice assumption. It has been proven that Nashian free choice violations and locality violations are equivalent resources in Bell experiments [38]. That result strengthens our proposal to investigate further quantum theory from a perspective where the free choice assumption is weakened. Let us also further observe that previously established constraints of non-extensibility of quantum theory to improved predictive power theories [32] do not hold in non-Nashian framework.

Finally, we note that the conclusions of our reasoning are independent on the exact form of agents' payoffs. Hence, in this work, we left them open. However, if the payoffs were set, a fixed point equilibrium might be found in a procedure resembling the considerations revoking to the least action principle. We would consider alternate histories which can be seen as perturbations of the examined history and by studying their properties, we would obtain the actual history that is Pareto-optimal for all agents.

## 4 Methods

*Perfect Prediction Equilibrium framework.* At this point, we present a brief description of the game theoretic notion of Perfect Prediction Equilibrium, first introduced in [18]. It is an equilibrium concept applicable to finite dynamic games with perfect information that allows to incorporate non-Nashian free choice into game theory. For dynamic games with strict preferences, the Perfect Prediction Equilibrium always exists. Additionally to that, it is unique and Pareto-optimal. We illustrate the inner-workings of the elimination algorithm used to solve for Perfect Prediction Equilibrium by first introducing two high-level principles it utilizes and then by showing a step-by-step procedure how it operates on a concrete example of a $\Gamma -$ game.

*first principle.* Consider one of the possible outcomes $o$, and an agent $p$ making a decision parameterized as $D$ on the path to $o$. The path from $D$ to $o$ is called a causal bridge. If it is so that, had the player made a different decision at $D$ than the one leading to $o$, it would guarantee them a higher payoff no matter what choice other players in future make, then the outcome $o$ is logically impossible (ad-absurdum proof) and discarded. This preemption process is said, in the original publication [18], to break the causal bridge to $o$ because its anticipation would not cause it. Figure 8a provides a schematic view of how this principle is applied in practice. Let us make clear that there is no causal effect on the past involved. The reasoning that agents use is purely based on counterfactual statements (for more details please refer to Sect. 2.2). Any agent playing after the agent $p$ would not have chosen the path ultimately leading to $o$ because agent $p$ would have anticipated that and hence would have deviated.

*second principle.* Consider all the outcomes that have not been eliminated yet. Note that all the discarded outcomes are no longer taken into account. If a player $p$ decides between two decisions $d_1$ and $d_2$, and all the payoffs connected to the outcomes following $d_1$ are higher than the maximum payoff for the outcomes subsequent to $d_2$, then it is rational for this agent to pick $d_1$. We visualize this concept by means of Fig. 8b.

As an example, let us consider a dynamic game without chance moves, with perfect information played between Alice and Bob who have strict preferences. We call this specific example $\Gamma -$ game (see Fig. 9). This particular game does not posses any physical or game-theoretic interpretation. It was created solely for pedagogical reasons of demonstrating how we solve a game for the equilibrium employing two introduced aforementioned principles. In the diagram (see Fig. 9), we see what are the consecutive steps in the process of solving for Perfect Prediction Equilibrium. The sequential steps are labeled by integer indices and are annotated with a description whether they use the first or the second principle. For more detailed description, please refer to [18].

*General spacetime positions* The algorithm for finding an equilibrium with non-Nashian free choice assumption also exists generally for any dynamic games with imperfect information, and in particular for spacetime games, assuming payoffs are in generic position (no ties between payoffs for the same agent). The generalization of the Perfect Prediction Equilibrium concept is called Perfectly Transparent Equilibrium [21].

*Spacetime games structure.* The additional objects needed to provide a structure for the outcomes of a spacetime game are as follows:

- a set $\mathcal{H}$ of decision nodes, each node $n \in H$ is associated with a player $\rho(n) \in \mathcal{P}$,
- a set of actions $\mathcal{A}$, where each decision node $N \in \mathcal{H}$ is associated with a subset of actions $\chi(N) \subseteq \mathcal{A}$ that players $\rho(N)$ can pick from.

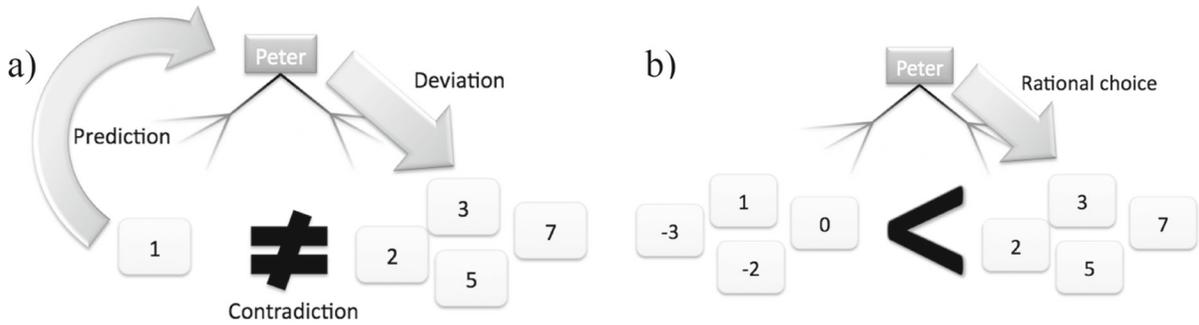

**Fig. 8** The two principles used in the Perfect Prediction Equilibrium framework, as they are used in practice. The principles were introduced in [18] and the pictures are also from this paper. The first principle (preemption) is shown on subfigure **a**. If Peter predicts he will get a payoff of 1 by going to the left, while *all* outcomes on the right are strictly greater than 1, then this is in direct contradiction with his rationality (*reductio ad absurdum*). Thus, the outcome giving Peter a payoff of 1 is a logical impossibility and is permanently discarded in the algorithm. The second principle (rational choice) is shown on subfigure **b**. If Peter sees that all outcomes on the right give him a higher payoff that any outcome on the left, then the rational decision is for Peter to go to the right for his next move. Image source: [18]

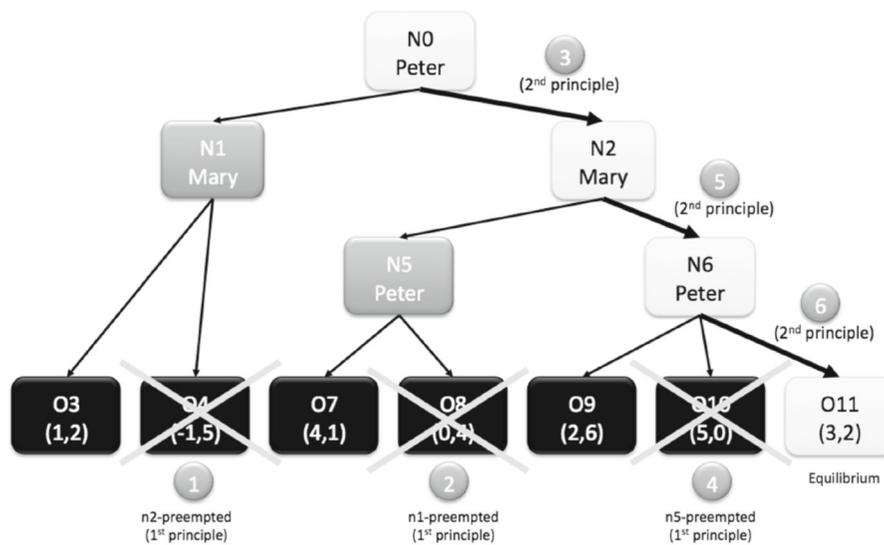

**Fig. 9** The non-Nashian resolution of a more complex game called the $\Gamma-$ game in the Perfect Prediction Equilibrium paper. It is a dynamic game with two players: Alice and Bob. To solve for the equilibrium, a process of forward induction utilizing the first and the second principle shown on Fig. 8 is implemented. The consecutive steps of the algorithms are indicated with round indexed tokens: outcome 4 is eliminated by virtue of the first principle (all outcomes on the right of node 0 are strictly greater than -1 for Peter). Then, outcome 8 is eliminated by virtue of the first principle (all remaining outcomes on the left of node 0 are strictly greater than 0 for Peter). Then, all outcomes on the right of node 0 are strictly greater than those on the left, so that Peter moves to node 2 by virtue of the second principle. Then, outcome 10 is eliminated by virtue of the first principle (all remaining outcomes on the left of node 2 are strictly greater than 0 for Mary). Then, Mary moves to node 6 by virtue of the second principle, and Peter moves to outcome 11 also by virtue of the second principle. The Perfect Prediction Equilibrium for the $\Gamma-$ game is $O_{11} = (3, 2)$. Image source: [18]

- A relation $\mathcal{R}$ that forms a directed acyclic graph over the nodes in $\mathcal{H}$.
- A label function $\sigma$ mapping each edge $(N, M) \in \mathcal{R}$ to an action $\sigma(N, M) \in \chi(N)$ available to $\rho(N)$ at that node.

The relation $\mathcal{R}$ constitutes the causal dependency graph of the spacetime game based on the Minkowski locations of the decisions. The label function $\sigma$ is referred to as the contingency coordinate system, which tells under which conditions on past choices a future decision takes or does not take place.

The way the game is played is that agents at the source nodes (with no previous nodes) make their decisions. Then, the decision-making propagates down the directed acyclic graph following the arrows. However, a decision is only made at a node (it is said to be active) if all its parents are active, and the decisions at the parents match the label on the edge connecting to that parent.

Specifically, a possible outcome $z$ is a partial function from $\mathcal{H}$ to $\mathcal{A}$ so that two conditions apply:

First, all root decision nodes have an assignment:

$$\forall N \in H, (\nexists M \in H \text{ s.t. } (M, N) \in R) \implies z(N) \in \chi(N)$$

Second, non-root decision nodes have an assignment or not depending on whether the assignment of their parent decision nodes matches the label on the edge:

$$\forall (N, M) \in R, z(N) = \sigma(N, M) \iff z(M) \in \chi(M)$$

$\mathcal{Z}$ is then defined as the set of all outcomes $z$ that fulfill the above two constraints. The payoffs are then defined on $\mathcal{Z}$.

**Acknowledgements**  We thank Prof. Gustavo Alonso for financing Michal Baczyk's assistant position in 2020 and 2021 during his Physics Master's studies at ETH Zürich. We thank Felipe Sulser for contributing code to automatically generate games and solve their equilibrium, as well as for generating large-scale datasets, during his Master's thesis. We also thank Ramon Gomm for having implemented, in his Bachelor's thesis, a resolution engine for the Perfectly Transparent Equilibrium as described in the corresponding technical report [21], and which is linked in the code availability section. Figures 1 to 7 were drawn by Michal Baczyk. Figures 8 and 9 are taken from the paper [18] co-authored by Ghislain Fourny. We thank Michal Friedman for proof-reading the paper.

**Author contributions**  The authors contributed equally to this paper. GF designed the formal model representing quantum experiments as games in Minkowski spacetime, contributed the non-Nashian solution concepts on various types of games, wrote a significant part of the paper. MB worked with GF as a research assistant and was involved in design discussions, wrote a significant part of the paper, prepared the pictures, structured the paper, prepared the paper for its final submission.

**Funding**  Open access funding provided by Swiss Federal Institute of Technology Zurich.

**Data availability**  A dataset with 200 million strategic games (i.e., spacelike-separated decisions) together with a Nash equilibrium resolution and a Perfectly Transparent Equilibrium resolution is publicly available [39]. An implementation in Java of the non-Nashian resolution (Perfectly Transparent Equilibrium) of any spacetime game is available at https://github.com/ghislainfourny/perfect-prediction-engine

**Declarations**

**Conflict of interest**  On behalf of all authors, the corresponding author states that there is no conflict of interest.